\documentclass[fleqn,usenatbib]{mnras}

\usepackage{newtxtext,newtxmath}

\usepackage[T1]{fontenc}

\DeclareRobustCommand{\VAN}[3]{#2}
\let\VANthebibliography\thebibliography
\def\thebibliography{\DeclareRobustCommand{\VAN}[3]{##3}\VANthebibliography}

\usepackage{graphicx}	
\usepackage{amsmath}	
\usepackage{orcidlink}
\usepackage[table]{xcolor}

\title[Environmental Dependency of MQG Assembly]{The Role of Environment in Shaping the Assembly Timelines of Massive Quiescent Galaxies at Cosmic Noon}

\author[Gao et al.]{
Tianmu Gao,$^{\orcidlink{0000-0002-1158-6372}1,2}$\thanks{E-mail: tianmu.gao@anu.edu.au}
J. Trevor Mendel,$^{\orcidlink{0000-0002-6327-9147}1,2}$
Letizia Bugiani,$^{\orcidlink{0009-0008-7295-3288}3,4}$
Minjung Park,$^{\orcidlink{0000-0002-8435-9402}5,6}$
Sirio Belli,$^{\orcidlink{0000-0002-5615-6018}3}$
\newauthor Joel Leja,$^{\orcidlink{0000-0001-6755-1315}7,8,9}$
Claudia del P. Lagos,$^{\orcidlink{0000-0003-3021-8564}10,2}$
Amir H. Khoram,$^{\orcidlink{0009-0009-6563-282X}3,4}$
Rebecca L. Davies,$^{\orcidlink{0000-0002-3324-4824}11}$
Razieh Emami,$^{\orcidlink{0000-0002-2791-5011}12}$
\newauthor Gabriel Maheson,$^{\orcidlink{0009-0005-8978-8598}3}$
Anita Zanella,$^{\orcidlink{0000-0001-8600-7008}4}$
Benjamin P. Collins$^{\orcidlink{0009-0005-9328-6283}3,13}$
\\
$^{1}$Research School of Astronomy and Astrophysics, Australian National University, Weston Creek, ACT 2611, Australia\\
$^{2}$ARC Centre of Excellence for All Sky Astrophysics in 3 Dimensions (ASTRO 3D), Australia\\
$^{3}$Dipartimento di Fisica e Astronomia, Universit\`a di Bologna, Via Gobetti 93/2, I-40129, Bologna, Italy\\
$^{4}$INAF, Astrophysics and Space Science Observatory Bologna, Via P. Gobetti 93/3, I-40129 Bologna, Italy\\
$^{5}$Kavli Institute for Cosmology, University of Cambridge, Madingley Road, Cambridge, CB3 0HA, UK\\
$^{6}$Cavendish Laboratory - Astrophysics Group, University of Cambridge, 19 JJ Thomson Avenue, Cambridge, CB3 0HE, UK\\
$^{7}$Department of Astronomy \& Astrophysics, The Pennsylvania State University, University Park, PA 16802, USA\\
$^{8}$Institute for Computational \& Data Sciences, The Pennsylvania State University, University Park, PA 16802, USA\\
$^{9}$Institute for Gravitation and the Cosmos, The Pennsylvania State University, University Park, PA 16802, USA\\
$^{10}$International Centre for Radio Astronomy Research, The University of Western Australia, 35 Stirling Highway, Crawley, WA 6009, Australia\\
$^{11}$Centre for Astrophysics and Supercomputing, Swinburne University of Technology, John Street Hawthorn, 3122, VIC, Australia\\
$^{12}$Center for Astrophysics $\vert$ Harvard \& Smithsonian, 60 Garden Street, Cambridge, MA 02138, USA\\
$^{13}$Departamento de Astronomía, Universidad de Chile, Camino El Observatorio 1515, Las Condes, Santiago, Chile
}

\date{Accepted XXX. Received YYY; in original form ZZZ}

\pubyear{\the\year{}}

\begin{document}
\label{firstpage}
\pagerange{\pageref{firstpage}--\pageref{lastpage}}
\maketitle

\begin{abstract}
\emph{JWST} observations of massive quiescent galaxies (MQGs) at $z\sim2$ have revealed diverse star formation histories (SFHs). In this letter, we investigate whether these variations are linked to the environment in which MQGs reside, as quantified by the host dark matter halo mass. We identify 12 MQGs from a representative sample of galaxies at $1.6<z<3.5$ obtained by the Blue Jay survey, a \emph{JWST} Cycle 1 program, and infer their SFHs via modelling of the deep NIRSpec medium-resolution spectra along with multi-band photometry. By correlating the formation and quenching epochs of individual MQGs with their respective host halo masses from the 3D-HST survey, we find that after accounting for the redshift range of the sample ($1.8<z<2.5$), MQGs in more massive haloes formed at significantly earlier cosmic epochs. Meanwhile, although these galaxies also quenched earlier overall, the scatter in quenching epochs across similar halo masses indicates that the quenching process likely involves other contributing factors and is not solely determined by halo mass.

\end{abstract}

\begin{keywords}
galaxies: formation -- galaxies: evolution -- galaxies: haloes
\end{keywords}



\section{Introduction}

The physical processes responsible for shutting down star formation in galaxies (i.e., quenching) have long been a major research focus \cite[for a review, see e.g.,][]{2018Man, 2026DeLucia, 2026Whitaker, 2026BluckP2, 2026BluckP1}, especially since \emph{JWST} revealed an overabundance of massive ($M_{\star}\gtrsim10^{10}M_{\odot}$) quiescent galaxies (MQGs) at $z>3$ \cite[e.g.,][]{2023Carnall,2023Valentino,2024Nanayakkara,2024Long,2024Alberts,2025BakerA,2025BakerB,2025Zhang,2025Stevenson}. A variety of mechanisms operating within galaxies have been proposed to contribute to their quenching, including feedback from stars and active galactic nuclei (AGN) \cite[e.g.,][]{2006Croton,2012Hopkins}, disk instability \cite[e.g.,][]{2014Dekel,2015Zolotov} and morphological quenching \cite[e.g.,][]{2009Martig}. Moreover, since early-type, passive galaxies in the local Universe are preferentially found in dense environments \cite[e.g.,][]{1980Dressler}, interactions with neighbouring galaxies and host dark matter haloes are also thought to influence their star-formation activity. Possible environmental mechanisms include galaxy mergers \citep[e.g.,][]{2001Genzel,2013Hopkins}, tidal encounters \citep[e.g.,][]{1981Farouki,1996Moore}, gas stripping \citep[e.g.,][]{1972Gunn,2001Vollmer} and the halt of gas supply \citep[e.g.,][]{1980Larson, 2015Peng}. Observational studies of low-redshift galaxies suggest that massive galaxies experience both rapid formation and quenching at earlier cosmic epochs, while star formation in low-mass galaxies proceeds over more extended timescales \citep[so-called "downsizing"; e.g.,][]{1996Cowie,2005Thomas,2006Bundy,2006Cimatti,2008Perez-Gonzalez,2010Thomas}. That said, the dominant quenching mechanisms at different redshifts, and the interplay among them, remain to be explored.

At $z\sim$1--3, the epoch known as "cosmic noon", \emph{JWST} observations have revealed a connection between the quenching of MQGs and multiphase AGN-driven outflows \citep{2024Davies,2024Scholtz,2024Belli,2024Park,2024DEugenio, 2025Bugiani}. Meanwhile, the role of environment is still unclear. Case studies using deep spectroscopy have revealed low-mass ($M_{\star}\sim10^{9}M_{\odot}$) quiescent galaxies at $z\sim2$ that reside in overdensities and show positive color gradients, indicative of environment-driven, outside-in quenching \citep{2023Sandles,2026Ito}. In contrast, a study of a protocluster at $z\sim2.6$ using photometric data from the UNCOVER \citep{2024Bezanson} and MegaScience \citep{2024Suess} surveys finds no significant differences in star formation histories (SFHs) between cluster members and field galaxies with $M_{\star}\lesssim10^{9}M_{\odot}$ \citep{2025Pan}. For massive galaxies, environmental effects are often difficult to disentangle from internal processes. Nevertheless, mergers are thought to facilitate quenching by driving starburst and AGN feedback \citep[e.g.,][]{2008HopkinsA,2014Toft}, and by enhancing turbulence \citep[e.g.,][]{2025Suess,2026Khoram}. In cosmological simulations, while the host haloes of MQGs are consistently predicted to undergo rapid early growth, different models predict distinct quenching epochs, thereby yielding varying contrasts in environment between these galaxies and their star-forming counterparts when star formation ceases. \citep[e.g.,][]{2023Hartley,2024Kurinchi-Vendhan,2025Kimmig,2025DeLuciaB,2025Chandro,2026Chittenden}.

Such diverse formation and quenching histories of high-redshift MQGs in simulations are likely linked to different baryonic implementations \citep[e.g., AGN feedback models;][]{2025Lagos}. SFHs inferred from rest-frame optical stellar continua, though subject to priors in spectral modelling, can therefore provide powerful constraints on their evolutionary pathways and the underlying physical mechanisms. Using deep medium-resolution ($R\sim1000$) \emph{JWST}/NIRSpec spectroscopy \citep{2022Jakobsen}, \cite{2024Park} derived SFHs for 14 MQGs at $z\sim2$ using data from the Blue Jay survey \citep{2025Belli}, showing that these galaxies follow three distinct formation pathways: old galaxies that formed and quenched at early epochs; galaxies with more extended SFHs that recently underwent rapid quenching; and galaxies that both formed and quenched recently following a major starburst. In this letter, we explore the connection between the formation pathways identified by \cite{2024Park} and the environment, as quantified by the host dark matter halo mass. We describe the Blue Jay survey, spectral modelling, and halo mass measurements in Section \ref{sec:data}. In Section \ref{sec:results}, we present the correlation between the formation/quenching epochs of the galaxies and their host halo masses. We discuss the implications of our findings in Section \ref{sec:discussion}. Throughout this work, we assume a flat $\Lambda$CDM cosmology with $H_0 = 70~\mathrm{km\,s^{-1}\,Mpc^{-1}}$, $\Omega_{\mathrm{m}} = 0.28$, and $\Omega_{\Lambda} = 0.72$.

\section{Data}\label{sec:data}
\subsection{The Blue Jay Survey}

The Blue Jay survey is a \emph{JWST} Cycle 1 GO program (PID: 1810; PI: Belli) designed to study both stars and gas in galaxies at cosmic noon through ultra-deep spectroscopy with the NIRSpec micro-shutter assembly \citep[MSA;][]{2022Ferruit}. The sample consists of 149 galaxies with $9<\log(M_{\star}/M_{\odot})<11.5$ at $1.7<z<3.5$, observed in two MSA pointings in the COSMOS field. To obtain full coverage of the rest-frame optical wavelength range, all three medium-resolution gratings (G140M, G235M, and G395M) were used, with exposure times of 13, 3.2, and 1.6 hours, respectively. For more details about the sample selection, observations and data reduction, we refer the reader to the overview paper of the program \citep{2025Belli}. Such deep observations are sufficient to achieve robust detections of the stellar continuum ($S/N\approx10$) for the MQGs in our sample, which are the primary targets of this work.

Stellar masses and SFHs for Blue Jay galaxies were derived using \textsc{Prospector} \citep{2021Johnson}, adopting a broadly similar model setup and priors to \citet{2024Park}. In this work, we use updated modelling described by Bugiani et al. (in prep.). We briefly summarise the relevant details below, but refer the reader to \citet{2024Park} and Bugiani et al. for more details.

Compared with \citet{2024Park}, Bugiani et al. use two different sets of photometric measurements, both derived using \textit{JWST}/NIRCam imaging from PRIMER (PID: 1837; PI: Dunlop), and \emph{HST}/ACS and WFC3 imaging from 3D-HST/CANDELS \citep{2014Skelton}. The first is extracted within an elliptical aperture defined using the NIRCam/F444W segmentation map and is taken to represent the total source. The second is extracted from a region defined by the MSA shutter footprint and is used for fits that include both photometry and spectroscopy. We refer to these as ``source'' and ``box'' photometric fits hereafter. A more detailed description of the photometric and spectral extraction is provided in \citet{2025Belli}.

Galaxy spectral energy distributions (SEDs) over 0.6\text{--}5 $\mu\mathrm{m}$ were modelled with \textsc{Prospector}, adopting a non-parametric SFH with 14 time bins. The youngest bin spans 0\text{--}30 Myr, followed by bins that are equally spaced in logarithmic space up to 0.8\,$t_{\rm univ}$, where $t_{\rm univ}$ is the age of the Universe at the observed redshift. The two oldest bins span 0.8\text{--}0.9\,$t_{\rm univ}$ and 0.9\text{--}1.0\,$t_{\rm univ}$, respectively. We adopt MIST isochrones \citep{2016Choi} and the MILES stellar spectral library \citep{2010Vazdekis, 2015Vazdekis}, and assume a Chabrier initial mass function (IMF; \citealt{2005Chabrier}) for all fits. Dust attenuation is included using a two-component model following \citet{2000Charlot}. As in \citet{2024Park}, we consider two different SFH priors, a ``continuity'' prior that favours a smooth SFH \citep{2019Leja}, and a ``bursty'' prior that allows for greater variability \citep{2022Tacchella}. For box fits, we fit spectra over the rest-frame wavelength range of 4000\text{--}6700 \AA, and include a 10th-order polynomial to account for imperfect spectral flux calibration and slit losses. The Na I D absorption feature at $\sim5900$\,\AA\ was masked, as it may originate not only from the stellar population but also from intervening neutral interstellar medium. Unless otherwise stated, we refer to results using the continuity prior, but note cases where the choice of SFH prior qualitatively changes the results.

The star-formation rate (SFR) of each galaxy is calculated by averaging the SFH over the most recent 30 Myr, corresponding to the youngest SFH bin. We identify quiescent galaxies by requiring that $\mathrm{sSFR} < 0.2/t_{\mathrm{univ}}$ \citep{2008Franx}, where sSFR is the specific star-formation rate ($\mathrm{sSFR}=\mathrm{SFR}/M_{\star}$), in both the source and box fits. This selection yields a sample of 12 MQGs, which differs slightly from the sample in \cite{2024Park}, primarily due to the use of the box aperture described above.\footnote{Specifically, COSMOS-9871, 19572, and 21541 are now classified as quiescent, while COSMOS-7549, 8013, 10400, 18668, and 21477 are no longer classified as quiescent.}

\subsection{Halo Mass Measurements}

11 out of the 12 Blue Jay MQGs have host halo mass estimates from \cite{2017Fossati}, based on data from the 3D-HST survey \citep{2012Brammer,2016Momcheva}. The 3D-HST survey extracted \emph{HST}/WFC3 G141 slitless spectra ($R\sim100$) for objects brighter than 26 mag in the F140W band (hereafter $JH_{140}$) in the CANDELS fields \citep{2011Grogin,2011Koekemoer}. In combination with extensive ancillary photometric data \citep{2014Skelton}, it provides redshift measurements with typical uncertainties of $\Delta z/(1+z)\sim0.003$ \citep{2016Bezanson}. \cite{2017Fossati} calculated the projected environmental density within a 0.75 Mpc aperture for all sources with $JH_{140}<24$ mag and $0.5<z<3$, whose redshift fits were visually inspected \citep{2016Momcheva}. Mock galaxies from the Munich semi-analytic model \citep{2015Henriques}, reproducing the 3D-HST sample selection and redshift accuracy were matched to the observed galaxies based on redshift, environmental density, stellar mass, and mass rank. Each 3D-HST galaxy was then assigned a probability of being a central or a satellite galaxy, along with a host halo mass probability distribution function (PDF) for each case, derived from the relative fractions and the halo masses of the matched central and satellite mock galaxies, respectively. Further details on the central and satellite assignment and halo mass inference, including how the approach statistically accounts for the impact of grism redshift uncertainties on halo membership identification, can be found in \cite{2015Fossati} and \cite{2017Fossati}. COSMOS-10339 is fainter than the $JH_{140}=24$ mag limit and is excluded from subsequent analyses. We note that for most galaxies in our sample, the probability of being a central galaxy exceeds 0.9, except for COSMOS-8469 ($P_{\rm cen}=0.658$) and 10592 ($P_{\rm cen}=0.589$). Since our results are not significantly affected by this choice, we treat all galaxies in our sample as central galaxies and adopt the corresponding halo masses.

\section{Results}\label{sec:results}

We measure the formation and quenching epochs from the SFHs inferred from the box fits, following the definitions adopted in \cite{2024Park}. The formation epoch is defined as the time when half of the final stellar mass has formed, and is characterised by the formation redshift $z_{\mathrm{form}}$ and the lookback time $t^{\mathrm{lbt}}_{\mathrm{form}}$ (measured in Gyr relative to the epoch of observation). For the quenching epoch, $z_{\mathrm{quench}}$ and $t^{\mathrm{lbt}}_{\mathrm{quench}}$ are defined based on the epoch when the sSFR falls below $\mathrm{sSFR} < 0.2/t_{\mathrm{univ}}(z)$. The uncertainties of these quantities are estimated from the 16th\text{--}84th percentile ranges of their posterior distributions derived from the inferred SFHs. We compare the
colours probing the rest-frame optical stellar continuum within the
box aperture with those of the remaining galaxy light and find only
small differences ($\lesssim 0.1$ mag) for all galaxies in our sample, suggesting that the stellar populations sampled by the spectra are broadly representative. According to the classification criteria of \cite{2024Park}, we identify 6 early-formed and early-quenched ($t^{\mathrm{lbt}}_{\mathrm{form}}>1.0$ Gyr, $t^{\mathrm{lbt}}_{\mathrm{quench}}>0.5$ Gyr), 4 early-formed and recently-quenched ($t^{\mathrm{lbt}}_{\mathrm{form}}>1.0$ Gyr, $t^{\mathrm{lbt}}_{\mathrm{quench}}\leq0.5$ Gyr), and 2 recently-formed and recently-quenched ($t^{\mathrm{lbt}}_{\mathrm{form}}\leq1.0$ Gyr, $t^{\mathrm{lbt}}_{\mathrm{quench}}\leq0.5$ Gyr) galaxies based on the continuity-prior fits. The properties of the galaxies in our sample are summarised in Table \ref{tab:sample}.

\begin{table*}
\renewcommand{\arraystretch}{1.2}
    \centering
    \caption{Properties of Blue Jay massive quiescent galaxies derived from \textsc{Prospector} fits adopting a continuity prior. Column 1: COSMOS ID. Column 2: Observed redshift. Column 3: (Survived) stellar mass derived from SED modelling using photometry measured within an elliptical aperture alone. Column 4: Formation redshift. Column 5: Formation epoch in lookback time (in Gyr). Column 6: Quenching redshift. Column 7: Quenching epoch in lookback time (in Gyr). Column 8: Host halo mass from \citet{2017Fossati}. Uncertainties quoted here correspond to the 16th to 84th percentiles of the posterior PDFs.}
    \begin{tabular}{ccccccccc}
        \hline
        & COSMOS ID & $z_{\mathrm{obs}}$ & $\log({M_{\star}/M_{\odot}})$ & $z_{\mathrm{form}}$ & $t^{\mathrm{lbt}}_{\mathrm{form}}$ & $z_{\mathrm{quench}}$ & $t^{\mathrm{lbt}}_{\mathrm{quench}}$ & $\log({M_{\mathrm{halo}}/M_{\odot}})$ \\
        \hline 
        \rowcolor{red!5} \multicolumn{9}{l}{Early-formed and Early-quenched} \\
        \hline
        1 & 8469 & 1.867 & $10.50^{+0.16}_{-0.89}$ & $5.14^{+2.96}_{-1.01}$ & $2.40^{+0.51}_{-0.35}$ & $2.21^{+0.22}_{-0.13}$ & $0.54^{+0.28}_{-0.19}$ & $12.12^{+0.28}_{-0.20}$ \\
        2 & 10128 & 1.851 & $11.25^{+0.02}_{-0.00}$ & $10.92^{+4.27}_{-4.69}$ & $3.16^{+0.15}_{-0.48}$ & $3.74^{+3.64}_{-0.00}$ & $1.89^{+0.97}_{-0.00}$ & $12.81^{+0.31}_{-0.34}$ \\ 
        3 & 10565 & 2.443 & $10.72^{+0.14}_{-0.03}$ & $5.17^{+4.29}_{-0.90}$ & $1.58^{+0.62}_{-0.30}$ & $3.17^{+0.50}_{-0.54}$ & $0.67^{+0.32}_{-0.46}$ & $12.39^{+0.24}_{-0.23}$ \\
        4 & 10592 & 1.800 & $11.18^{+0.04}_{-0.06}$ & $15.46^{+4.48}_{-8.02}$ & $3.41^{+0.08}_{-0.45}$ &  $7.27^{+4.68}_{-4.92}$ & $2.94^{+0.36}_{-2.10}$ & $12.71^{+0.36}_{-0.35}$ \\
        5 & 11494 & 2.090 & $11.74^{+0.03}_{-0.02}$ & $9.77^{+6.09}_{-3.15}$ & $2.69^{+0.24}_{-0.34}$ & $4.18^{+3.86}_{-1.47}$ & $1.70^{+0.84}_{-0.95}$ & $13.15^{+0.29}_{-0.27}$ \\
        6 & 16419 & 1.926 & $11.27^{+0.02}_{-0.02}$ & $16.34^{+3.52}_{-3.19}$ &  $3.21^{+0.06}_{-0.09}$ &  $7.62^{+4.90}_{-0.00}$ & $2.76^{+0.34}_{-0.00}$ & $13.02^{+0.28}_{-0.29}$ \\
       
        \hline 
        \rowcolor{green!5}\multicolumn{9}{l}{Early-formed and Recently-quenched} \\
        \hline
        7 & 9395 & 2.127 & $10.81^{+0.10}_{-0.06}$ & $3.83^{+0.72}_{-0.46}$ &  $1.48^{+0.32}_{-0.26}$ &  $2.52^{+0.24}_{-0.23}$ & $0.50^{+0.24}_{-0.28}$ & $12.36^{+0.28}_{-0.24}$ \\
        8 & 9871 & 2.045 & $11.63^{+0.13}_{-0.23}$ & $4.30^{+2.36}_{-1.25}$ &  $1.82^{+0.61}_{-0.71}$ &  $2.15^{+0.51}_{-0.10}$ &  $0.15^{+0.62}_{-0.15}$ & $13.09^{+0.30}_{-0.30}$ \\
        9 & 19572 & 1.868 & $11.03^{+0.06}_{-0.06}$ & $3.09^{+1.05}_{-0.44}$ & $1.44^{+0.61}_{-0.39}$ &  $1.95^{+0.13}_{-0.06}$ & $0.15^{+0.20}_{-0.11}$ & $12.55^{+0.28}_{-0.28}$ \\
        10 & 21541 & 2.476 & $10.90^{+0.03}_{-0.11}$ & $5.94^{+2.32}_{-0.70}$ & $1.72^{+0.34}_{-0.16}$ & $2.94^{+0.27}_{-0.00}$ & $0.45^{+0.21}_{-0.00}$ & $12.58^{+0.24}_{-0.25}$ \\
        \hline 
        \rowcolor{blue!5}\multicolumn{9}{l}{Recently-formed and Recently-quenched} \\
        \hline
        11 & 10339 & 2.363 & $10.41^{+0.16}_{-0.13}$ & $3.08^{+0.32}_{-0.13}$ &  $0.70^{+0.22}_{-0.11}$ &  $2.44^{+0.11}_{-0.04}$ &  $0.09^{+0.12}_{-0.05}$ & / \\
        12 & 11142 & 2.445 & $11.04^{+0.06}_{-0.06}$ & $3.00^{+0.08}_{-0.08}$ &  $0.54^{+0.06}_{-0.07}$ &  $2.57^{+0.07}_{-0.09}$ &  $0.14^{+0.07}_{-0.10}$ & $12.52^{+0.32}_{-0.24}$ \\
        \hline
    \end{tabular}

    \label{tab:sample}
\end{table*}

\begin{figure}
    \centering
    \includegraphics[width=\linewidth]{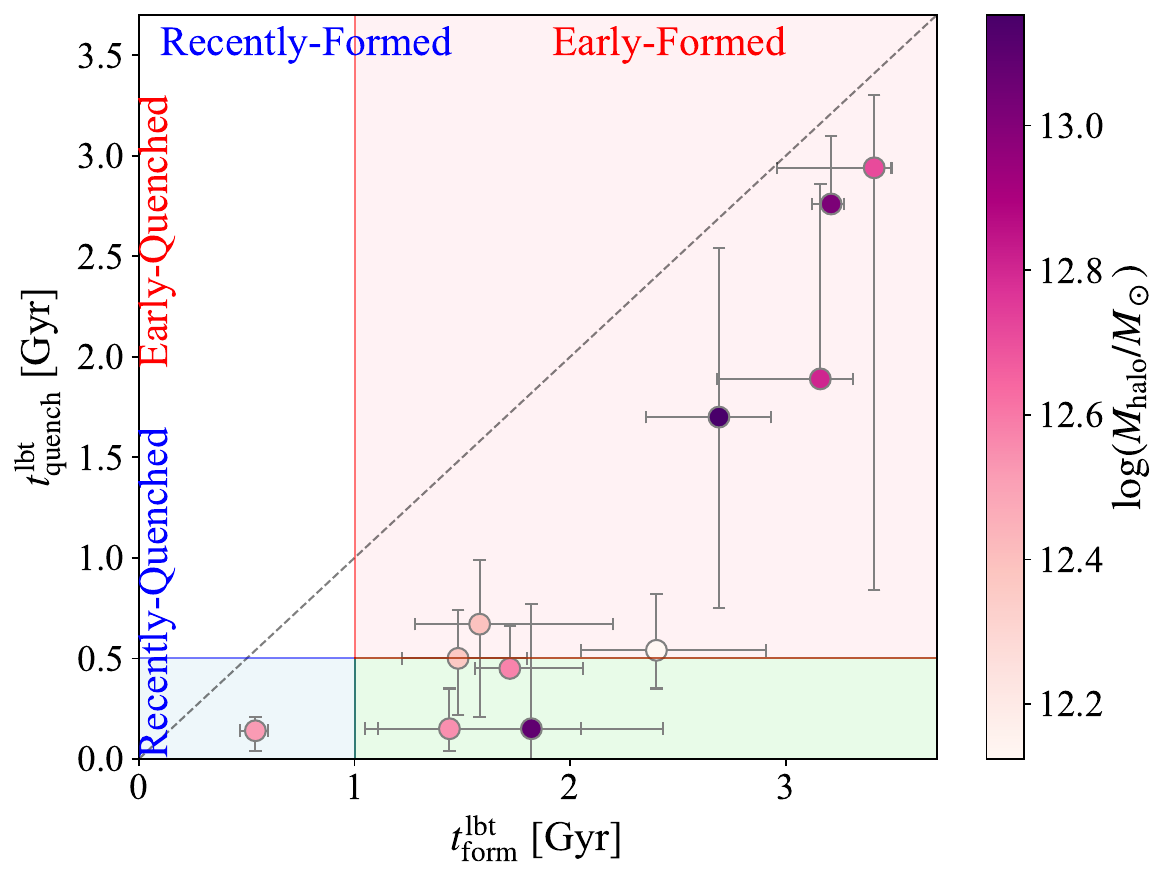}
    \caption{Distribution of Blue Jay MQGs in the plane of formation epoch ($t^{\mathrm{lbt}}_{\mathrm{form}}$) versus quenching epoch ($t^{\mathrm{lbt}}_{\mathrm{quench}}$), both in lookback time (Gyr). Galaxies are color-coded by host halo mass from \citet{2017Fossati}.}
    \label{fig:SFHvsEnv}
\end{figure}

In Figure \ref{fig:SFHvsEnv}, we show the distribution of the 11 Blue Jay MQGs (out of a total of 12) with available halo mass measurements, in the plane of formation ($t^{\mathrm{lbt}}_{\mathrm{form}}$) and quenching ($t^{\mathrm{lbt}}_{\mathrm{quench}}$) epochs, with each data point colour-coded by halo mass. We find that there is a halo mass dependence for both the formation and quenching times, with galaxies that formed and quenched at greater lookback times tending to reside in more massive haloes. To better quantify this trend, we divide our sample into two based on the median halo mass of the full sample, $\log(M_{\mathrm{halo}}/M_\odot)=12.58$. Galaxies in the high-mass halo group have a median $t^{\mathrm{lbt}}_{\mathrm{form}}=2.83^{+0.18}_{-0.22}$ Gyr, compared to $1.51^{+0.29}_{-0.22}$ Gyr for the low-mass halo group. We find a similar, though less significant, offset in quenching times, with $t^{\mathrm{lbt}}_{\mathrm{quench}}=1.73^{+0.51}_{-0.50}$ Gyr and $0.36^{+0.17}_{-0.16}$ Gyr for the high- and low-mass halo groups, respectively. 
Adopting the bursty prior shifts the inferred quenching times of some galaxies to larger values (i.e., earlier quenching), while the difference between the two halo-mass groups remains, with $t^{\mathrm{lbt}}_{\mathrm{form}}=2.98^{+0.15}_{-0.12}$ Gyr and $t^{\mathrm{lbt}}_{\mathrm{quench}}=2.68^{+0.20}_{-0.28}$ Gyr for the high-mass group, and $t^{\mathrm{lbt}}_{\mathrm{form}}=1.23^{+0.28}_{-0.20}$ Gyr and $t^{\mathrm{lbt}}_{\mathrm{quench}}=0.41^{+0.17}_{-0.08}$ Gyr for the low-mass group. We also verify that the qualitative results do not change when adopting the \citet{2024Park} MQG sample and measurements.

\begin{figure*}
    \centering

    \includegraphics[width=0.49\linewidth]{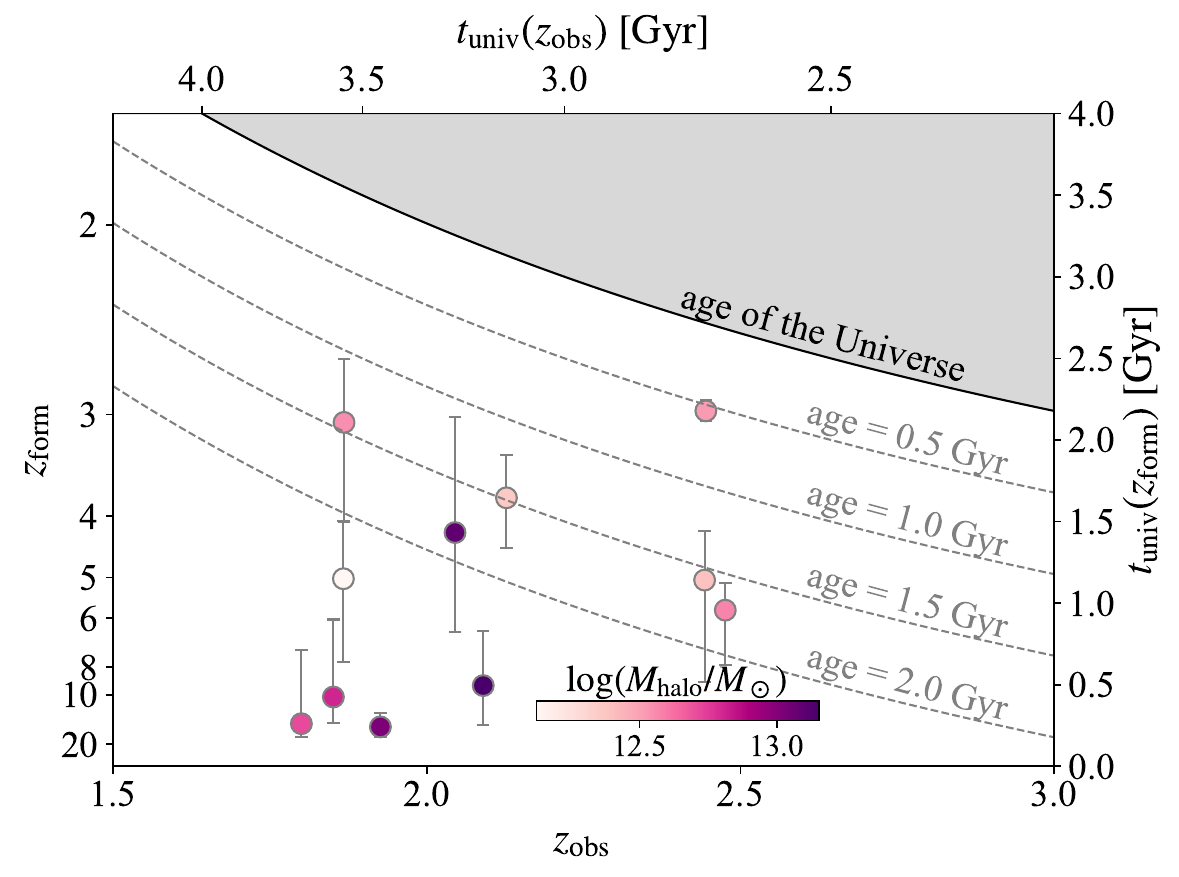}
    \hfill
    \includegraphics[width=0.49\linewidth]{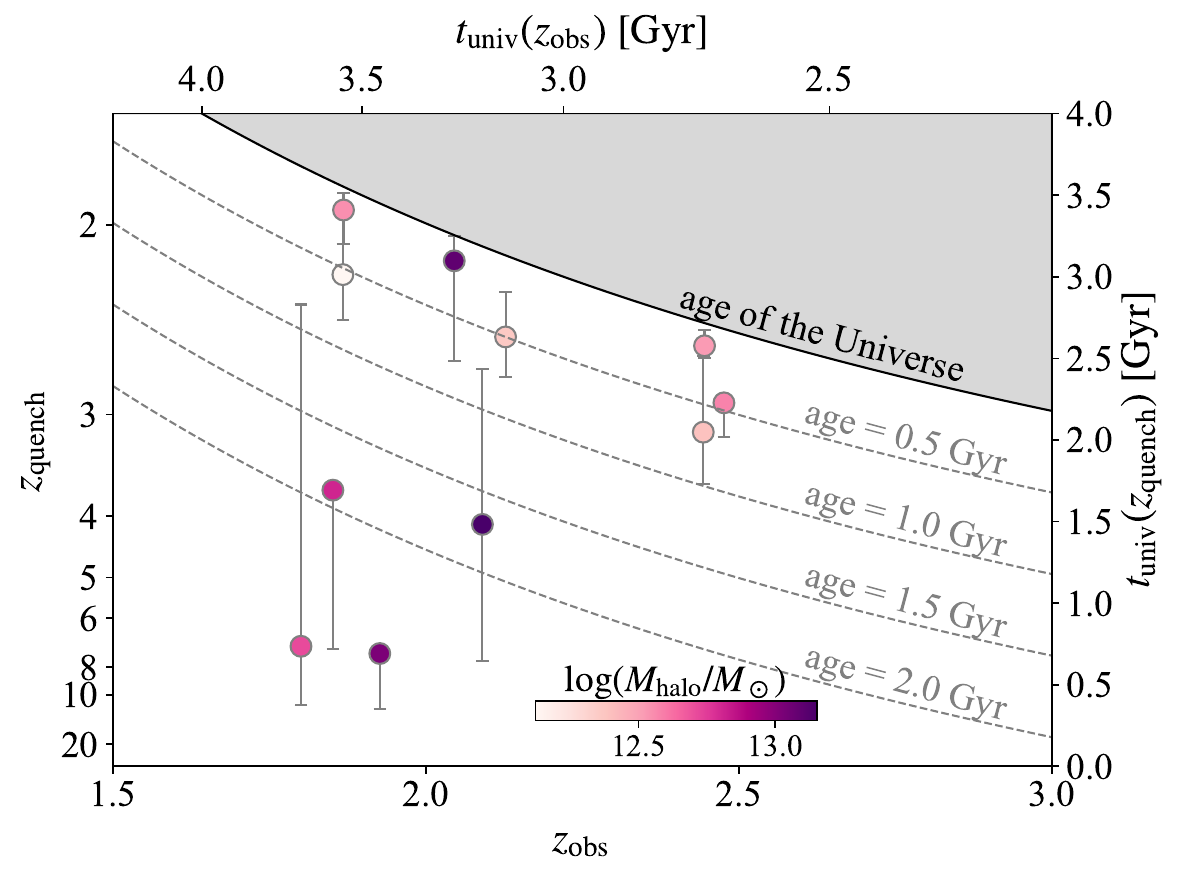}
    
    \caption{Formation redshift ($z_{\mathrm{form}}$, left) and quenching redshift ($z_{\mathrm{quench}}$, right) versus observed redshift ($z_{\mathrm{obs}}$) for the Blue Jay MQGs. Galaxies are color-coded by host halo mass from \citet{2017Fossati}. The dashed curves show constant lookback times of 0.5, 1.0, 1.5, and 2.0~Gyr relative to the observed epoch.}
    \label{fig:zobs_vs_zform_and_zquench}
\end{figure*}

As the galaxies in our sample span a redshift range of $z\approx$1.8--2.5, corresponding to a cosmic time interval of $\sim1$ Gyr, we examine their absolute formation and quenching epochs rather than lookback times. In Figure \ref{fig:zobs_vs_zform_and_zquench}, we plot the formation ($z_{\mathrm{form}}$; left panel) and quenching ($z_{\mathrm{quench}}$; right panel) redshifts against the observed redshifts, $z_{\mathrm{obs}}$, of all 11 galaxies, colour-coded by their host halo masses. Galaxies in more massive haloes also appear to have formed and quenched at higher median redshifts. Adopting the continuity prior, the high-mass group has median $z_{\mathrm{form}}=9.86^{+2.49}_{-2.27}$ and $z_{\mathrm{quench}} = 4.21^{+1.50}_{-0.84}$, compared with $z_{\mathrm{form}} = 3.90^{+0.58}_{-0.47}$ and $z_{\mathrm{quench}} = 2.47^{+0.11}_{-0.16}$ for the low-mass group. These differences are consistent with those seen in $t^{\mathrm{lbt}}_{\mathrm{form}}$ and $t^{\mathrm{lbt}}_{\mathrm{quench}}$, implying genuinely earlier median formation and quenching epochs for MQGs in more massive haloes.

Given the downsizing trend, we further assess the role of $M_{\mathrm{halo}}$ relative to $M_{\star}$ by calculating the Spearman correlation coefficients. Using the values based on the continuity-prior fits listed in Table \ref{tab:sample}, $M_{\mathrm{halo}}$ shows a positive correlation with $z_{\mathrm{form}}$ ($r=0.53$, $p=0.10$) and $z_{\mathrm{quench}}$ ($r=0.45$, $p=0.17$). Adopting the bursty prior strengthens both correlations, with $r=0.67$, $p=0.02$ for $z_{\mathrm{form}}$ and $r=0.65$, $p=0.03$ for $z_{\mathrm{quench}}$. Furthermore, when controlling for $M_{\star}$, a positive partial correlation between $M_{\mathrm{halo}}$ and $z_{\mathrm{form}}$ is also seen ($r=0.56$, $p=0.09$), with a weaker one for the bursty prior ($r=0.48$, $p=0.16$), though neither is statistically significant. However, $M_{\mathrm{halo}}$ and $M_{\star}$ are themselves strongly correlated ($r=0.95$, $p<0.001$). We therefore cannot robustly establish an independent role of $M_{\mathrm{halo}}$ in determining the formation epoch with the current sample. Meanwhile, the partial correlation between $M_{\mathrm{halo}}$ and $z_{\mathrm{quench}}$ is also positive but not statistically significant ($p\gtrsim0.2$), regardless of the choice of SFH prior.

Additionally, we note that more massive haloes are preferentially observed at lower redshifts. This may reflect a higher abundance of massive haloes at later epochs, as haloes observed at lower redshifts have had more time to accrete mass. We estimate the expected masses of all haloes at a common redshift of $z=1.80$ (the minimum redshift of our galaxy sample), following the prescription in equation 2 of \cite{2010Fakhouri}. Overall, the resulting changes are small: the largest corrections are comparable to the halo-mass measurement uncertainties, while those for the remaining galaxies are much smaller. This suggests that halo growth over the redshift range of our sample is unlikely to be the dominant factor of the observed trends. Nevertheless, the redshift range remains a source of uncertainty, and evolving the halo masses alone is insufficient for a fully self-consistent comparison. This further highlights the difficulty of robustly constraining the partial correlations with the current sample.

Beyond halo growth, the tendency for more massive haloes to be observed at lower redshifts may also arise from the limitations of using environmental density as a proxy for halo mass at higher redshifts. Specifically, in a flux-limited survey, the detectability of satellite galaxies decreases with redshift, compressing the dynamic range of the observed environmental density. This is supported by the observed neighbour counts for galaxies in our sample, which drop from 2\text{--}5 at $z\sim2$ to 0\text{--}2 at $z\sim2.5$. This reduced contrast may bias rank-based halo mass assignments, leading to underestimated halo masses in intrinsically dense environments and overestimated halo masses in sparse ones.

\section{Discussion}\label{sec:discussion}

\subsection{Dependence of the formation epoch on host halo mass}

Our finding that MQGs in more massive haloes tend to form earlier is consistent with previous work targeting early-formed quiescent galaxies at higher redshifts, which are often found in overdense regions \citep[e.g.,][]{2022McConachie,2026McConachie,2024Stawinski,2024Carnall,2025deGraaffA,2025Turner, 2026KakimotoA}. We note that although the studies discussed here generally characterise environment based on the distribution of surrounding galaxies rather than directly using host halo mass, such overdensities are often associated with massive haloes \citep{2012Haas, 2012Muldrew}. Potential evidence for the clustering of UV-bright galaxies at $z\gtrsim10$ also indicates that environment may already influence the formation of the first galaxies \citep[e.g.,][]{2023Tacchella,2024Carniani}. The role of massive haloes in early formation can be naturally understood as providing larger gas reservoirs and sustaining higher gas inflow rates \citep[e.g.,][]{2005Keres,2009DekelA,2025Hughes,2025Zhou}. Recent observations of $z\gtrsim4$ protoclusters have revealed their significant role in accelerating the early mass growth of their member galaxies \citep[e.g.,][]{2025Morishita,2025LiQiong,2025Lagache,2025WittenA,2025WittenB,2025Fudamoto, 2025Hatamnia}. Notably, these protoclusters are often found to host overdensities of dusty star-forming galaxies (DSFGs) \citep[e.g.,][]{2009Daddi,2018Oteo,2025WittenA,2025Lagache,2026Zeng}. The SFRs and stellar masses of DSFGs are consistent with the rapid growth phase inferred from the SFHs of MQGs observed at later epochs \citep[e.g.,][]{2020ForrestA,2020ForrestB, 2020Valentino, 2026Sun}, while their short gas-depletion timescales suggest that they could transition rapidly to quiescence \citep{2025Lagache}. Furthermore, these populations at different redshifts show broadly similar number densities, with $\sim3$\text{--}$6\times10^{-6}\,\mathrm{Mpc}^{-3}$ for DSFGs and a few $\times10^{-5}\,\mathrm{Mpc}^{-3}$ for both MQGs and UV-bright galaxies \citep{2026Sun,2026Zavala}. Some simulations also show that the progenitors of MQGs at $z\sim3$ exhibit SFRs and stellar masses consistent with DSFGs at $4<z<5$ \citep{2025Lagos}. Taken together, these findings hint at a potential evolutionary link between early-formed MQGs at $z\sim2$, DSFGs in protoclusters at $z\gtrsim4$, and UV-bright galaxies at $z\gtrsim10$ that exhibit clustering signatures.

A recently proposed ``feedback-free starburst'' (FFB) model suggests that, due to the high gas densities and low metallicities at early times, the free-fall time of gas clouds in massive dark matter haloes can be shorter than the timescale on which stellar feedback from winds and supernovae becomes effective \citep{2023Dekel,2023Renzini,2024Li}. This enables an efficient conversion of gas into stars, which may explain the early formation of those galaxies in our sample inferred to have assembled $\sim10^{10}\,M_{\odot}$ of stellar mass by $z_{\mathrm{form}}\gtrsim10$, although these estimates of early mass growth are subject to modelling assumptions and systematics.

\begin{figure}
    \centering
    \includegraphics[width=\linewidth]{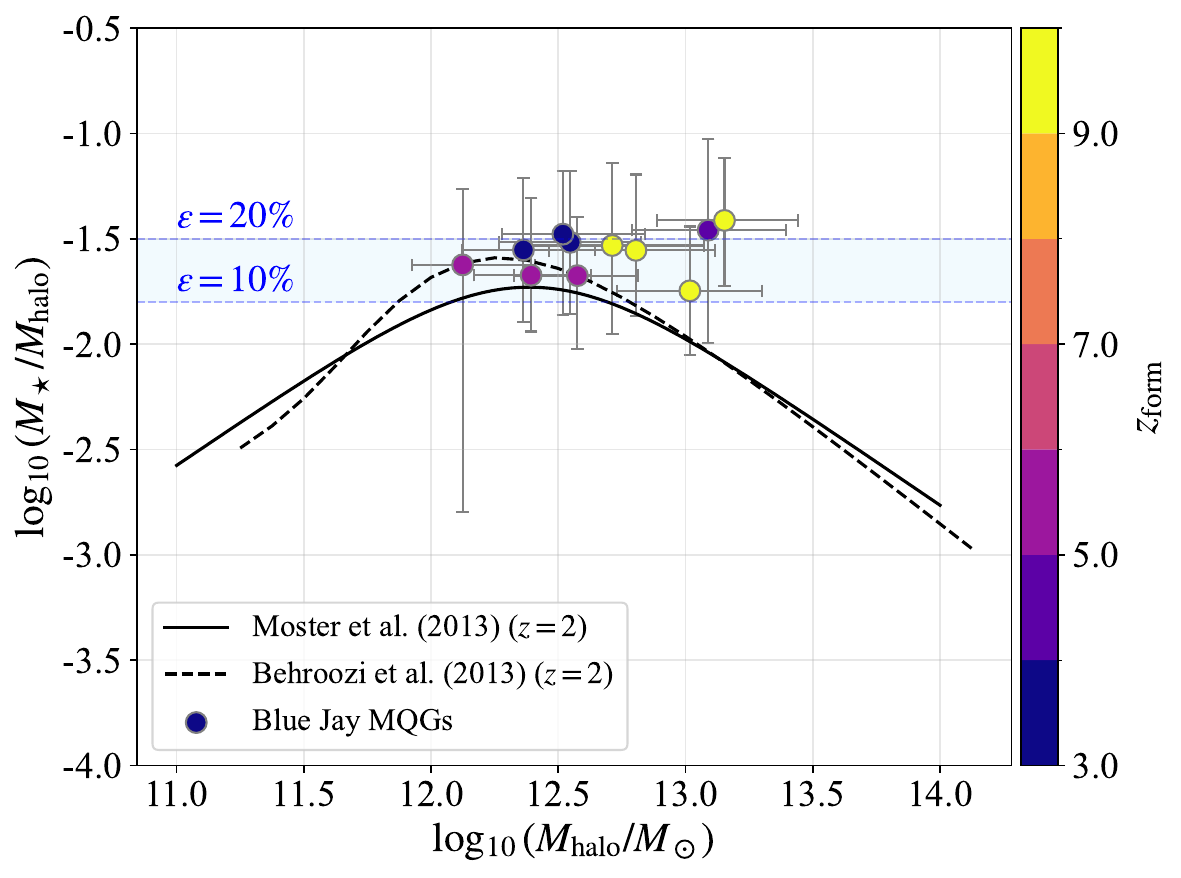}
    \caption{Stellar-to-halo mass ratio versus halo mass for the Blue Jay MQGs. Galaxies are colorcoded by $z_{\rm form}$. The solid and dashed lines represent the relations at $z=2$ from \citet{2013Moster} and \citet{2013Behroozi}, respectively. The shaded region indicates a baryon conversion efficiency $\epsilon$ between $10\%$ and $20\%$, where $M_{\star}=\epsilon f_{\rm b}M_{\rm halo}$ and $f_{\rm b}=0.158$ \citep{2020Planck}. Note that $\epsilon$ is derived based on the surviving stellar mass.}
    \label{fig:SHMR}
\end{figure}

We plot the stellar-to-halo mass relation (SHMR) in Figure \ref{fig:SHMR}. Assuming $M_{\star}=\epsilon f_{\rm b}M_{\rm halo}$ with a cosmic baryon fraction $f_{\rm b}=0.158$ \citep{2020Planck}, all galaxies in our sample exhibit a baryon conversion efficiency $\epsilon>10\%$, with some cases exceeding $20\%$, a level in line with that inferred for massive galaxies at higher redshifts \citep[e.g.,][]{2024Stawinski,2024Xiao,2025Antwi-Danso,2025deGraaffA}. We note that since $\epsilon$ is calculated here using the surviving stellar mass listed in Table \ref{tab:sample}, the total integrated star formation would yield even higher values. Furthermore, galaxies with $z_{\mathrm{form}}\gtrsim9$ lie significantly above both the \citet{2013Moster} and \citet{2013Behroozi} SHMRs at $z=2$, indicative of a highly efficient early star formation. Other studies have also found massive galaxies with very early formation epochs \citep[e.g.,][]{2024Glazebrook,2024Carnall,2026McConachie}, for which a substantial amount of stellar mass was already in place at early times, requiring baryon conversion efficiencies approaching the limit allowed within the $\Lambda$CDM framework \citep[e.g.,][]{2023Boylan-Kolchin}. It is possible that the overall SHMR at $z_{\mathrm{form}}\gtrsim9$ has a higher normalisation, which would make these galaxies ``normal'' \citep[e.g.,][]{2025Shuntov, 2025Paquereau}. However, future work is necessary to determine the comprehensive evolution of the SHMR in the early Universe.

What remains unclear is the contribution of \textit{ex situ} mass via mergers. Based on the \textsc{Flamingo} simulations, \cite{2025BakerA} finds that, for MQGs at $2<z<5$, the contribution from dry mergers with stellar mass ratios $>$ 1:10 appears to be negligible, with the majority of their stellar mass formed \textit{in situ}. In contrast, results from the \textsc{Colibre} simulations show that, by the time they quench, MQGs have undergone substantially more mergers than their star-forming counterparts, which accelerates their mass assembly \citep{2025Chandro}. Observational studies of galaxies at $z\gtrsim3$ also show varying estimates for the contribution of major mergers to total mass growth, which may arise from discrepancies in the estimated \textit{in situ} star formation \citep{2025Duan,2025Puskas} and uncertainties in the assumed merger timescales \citep{2021Romano}. For our sample, according to \cite{2024Park}, the early-formed and early-quenched galaxies could be interpreted as descendants of galaxies that quenched rapidly at early epochs, with their SFHs smoothed by subsequent minor mergers. On the other hand, for the early-formed and recently-quenched galaxies, the accretion of old stars prior to recent quenching could produce either a past burst or a flat inferred SFH. Nevertheless, while simulations suggest that the \textit{ex situ} mass fraction increases with halo mass \citep[e.g.,][]{2018Pillepich,2022Remus}, any signal of accretion of external stars or star formation triggered by gas-rich mergers is difficult to clearly distinguish from the overall inferred SFHs, especially at earlier epochs where the coarser time resolution in the modelling naturally smooths out bursts or merger signatures.

\subsection{Dependence of the quenching epoch on host halo mass}

We find a wide range of quenching epochs among galaxies residing in similarly massive haloes (right panel of Figure \ref{fig:zobs_vs_zform_and_zquench}), suggesting that quenching involves greater complexity than a single pathway determined solely by halo mass. The direct contribution of environmental mechanisms to the quenching of massive galaxies at high redshift appears to be minor \citep[e.g.,][]{2025Hatamnia}. More massive haloes may instead provide the physical conditions that facilitate internal quenching mechanisms, such as AGN feedback \citep[e.g.,][]{2025Perez-Gonzalez, 2026KakimotoB}, which many simulations suggest plays a dominant role in quenching high-redshift MQGs \citep[e.g.,][]{2023Hartley, 2024Lagos, 2025Lagos, 2024Kurinchi-Vendhan, 2025Weller, 2025Ni, 2025Chandro, 2026Chittenden}. Indeed, both quasars and radio galaxies are often found to trace overdense regions at high redshifts \citep[e.g.,][]{2007Venemans,2013Wylezalek,2016Overzier,2026WangF}.

In this context, one possible explanation for this diversity in quenching epochs is that AGN activity leading to rapid quenching can be triggered by episodes of gas compaction, which may occur at different times in different galaxies. Gas-rich mergers represent one major mechanism for driving such gas compaction, while also inducing morphological transformations, including the destruction of rotating discs and the formation of compact spheroids \citep[e.g.,][]{2008HopkinsB, 2009Hopkins}. On the other hand, internal processes such as violent disc instabilities can also funnel gas towards the galaxy centre while preserving the disc structure \citep[e.g.,][]{2009DekelB, 2014Dekel}. The timing of these compaction events can vary substantially among galaxies residing in similarly massive haloes. Further diversity in their quenching histories may arise from differences in how AGN feedback operates in individual galaxies, as suggested by results from simulations employing different AGN feedback models \citep{2025Weller,2025Lagos}. Such differences may involve both the mechanisms through which AGN inject energy into their host galaxies (e.g., via winds or jets) and how effectively this energy couples to the surrounding gas \citep[e.g., see the review of][]{2024Harrison}. In addition, the central concentration of cold gas that can fuel AGN activity may also induce intense starbursts, providing another channel for rapid quenching through accelerated gas depletion and enhanced stellar feedback \citep[e.g.,][]{2019Belli, 2026Skarbinski, 2026Davies}.

At larger scales, the accretion of cold gas onto galaxies can also be regulated as their host haloes grow, resulting in a more gradual process of quenching. While studies have shown that higher BH masses are required to trigger quenching in more massive haloes \citep{2022Bluck,2024Goubert,2025Lim}, the specific halo assembly history also appears to modulate the star formation timeline. Given that the masses of the haloes in our sample all exceed the critical mass of $\sim10^{12}{M_{\odot}}$, where virial shock heating becomes operative \citep{2003Birnboim,2005Keres,2006Dekel} and gas-to-star conversion efficiency peaks \citep{2010Behroozi,2013Behroozi,2013Moster}, they may have crossed this threshold at different cosmic epochs depending on their assembly histories. As a result, the efficient heating of infalling gas and the suppression of cold gas replenishment from the cosmic web may occur at different stages of their evolution, affecting star formation more gradually over extended timescales.

\section{Conclusions}\label{sec:conclusion}

In this letter, by combining the SFHs of 11 MQGs selected from the Blue Jay survey with the corresponding halo masses, we find tentative evidence that MQGs at cosmic noon residing in more massive dark matter haloes tend to form earlier than those in less massive ones. This may be linked to the large gas reservoirs available in early overdensities, which can fuel both rapid star formation and accelerated SMBH growth. While correspondingly earlier quenching is indeed observed, the scatter in quenching epochs across similar halo masses suggests diverse quenching pathways that may originate from different AGN feedback channels and halo growth histories.

Since the evolution of star formation activity in early galaxies can be strongly influenced by the cumulative energy injection from AGN, a representative sample of both quiescent and star-forming galaxies, complemented by robust measurements of their SFHs and BH masses, will be pivotal for refining different feedback models \citep[e.g.,][]{2025Lagos}. Meanwhile, detailed measurements of stellar chemical abundance patterns can now be obtained with ultra-deep NIRSpec medium-resolution spectroscopy \citep[e.g.,][]{2025Beverage}, providing complementary constraints on star-formation timescales. Furthermore, deep, wide-area \emph{JWST} surveys such as COSMOS-Web, which sample a wide range of environments across cosmic epochs \citep[e.g.,][]{2025Hatamnia}, will provide more accurate environmental density estimates and, in the near future, improved halo mass measurements for individual galaxies. These data will enable further investigation into how environment regulates star formation and SMBH growth, shedding light on the galaxy–halo connection.

\section*{Acknowledgements}

We would like to thank the anonymous referee for providing constructive comments, which have greatly improved the quality and clarity of this manuscript. This research was supported by the International Space Science Institute (ISSI) in Bern, through ISSI International Team project 24-602 "Multiphase Outflows in Galaxies at Cosmic Noon". The Blue Jay Survey is funded in part by STScI Grant JWST-GO-01810. TG, JTM, and CdPL acknowledge support from the Australian Research Council through Discovery Project DP210101945, funded by the Australian Government. SB, LB, and AHK are funded by ERC StG 101076080. MP acknowledges support from the Kavli Institute for Cosmology, Cambridge and the Kavli Foundation, and King's College, Cambridge. RLD is supported by the Australian Research Council through the Discovery Early Career Researcher Award (DECRA) Fellowship DE240100136 funded by the Australian Government. BPC acknowledges support from the EPEC Doctorate Scholarship for the financing of their postgraduate studies. This work is based on observations made with the NASA/ESA/CSA James Webb Space Telescope. The data were obtained from the Mikulski Archive for Space Telescopes at the Space Telescope Science Institute, which is operated by the Association of Universities for Research in Astronomy, Inc., under NASA contract NAS 5-03127 for JWST. These observations are associated with program GO 1810 and GO 1837. This work also makes use of observations taken by the 3D-HST Treasury Program (GO 12177 and 12328) with the NASA/ESA HST, which is operated by the Association of Universities for Research in Astronomy, Inc., under NASA contract NAS 5-26555. OpenAI ChatGPT and Google Gemini were used for minor grammar checking and language polishing. No results, interpretations, or conclusions were generated by these tools.
\section*{Data Availability}

The Blue Jay data, including ancillary \emph{HST} and \emph{JWST} photometry, are publicly available; see \citet{2025Belli} for details of the data release. The environmental properties of 3D-HST galaxies, including halo mass PDFs, are available at \url{http://dx.doi.org/10.5281/zenodo.168056}; see \citet{2017Fossati} for details.



\bibliographystyle{mnras}
\bibliography{example} 




\appendix

\section{Some extra material}


\bsp	
\label{lastpage}
\end{document}